\documentclass[11pt]{article}
\usepackage{amssymb,amsmath,amsthm,latexsym,amsfonts,amscd,dsfont}

\usepackage{a4wide,graphicx,tikz}
\usepackage{authblk}

\usetikzlibrary{arrows,shapes,shadows,fit,backgrounds}

\tikzstyle{trader} = [circle, draw, top color=white, bottom color=blue!30, draw=blue!50!black!100, drop shadow, minimum height=4em]
\tikzstyle{bank} = [rectangle, draw, top color=white, bottom color=red!20, draw=red!50!black!100, drop shadow, rounded corners, minimum height=3em, text width=4em, text centered]
\tikzstyle{market} = [rectangle, draw, top color=white, bottom color=green!20, draw=green!50!black!100, drop shadow, rounded corners, minimum height=3em, text width=4em, text centered]
\tikzstyle{background} = [rectangle,fill=gray!10, inner sep=0.2cm, rounded corners=5mm]
\tikzstyle{line} = [draw, latex'-latex']
\tikzstyle{from} = [draw, latex'-]
\tikzstyle{to} = [draw, -latex']

\newcommand{\url}[1]{{\tt \small #1}}

\title{Illustrating a problem in the self-financing condition in two 2010-2011 papers on funding, collateral and discounting\thanks{We are grateful to Daniele Perini for interesting suggestions and discussion. Given the specific moment and the current turmoil affecting the top management of some important banks following the LIBOR fixing scandal, see for example \cite{bibscandals},
and given that the authors of the articles we are scrutinizing are all affiliated with one of said banks, we feel it is important to point out that this is not meant as a criticism to the quantitative teams of said bank. This is simply a technical paper addressing a problem in the above published articles and has nothing to do with the affiliation of such articles' authors, or with the articles publisher. Finally, this article reflects only the authors opinion and in no way represents the authors employers, present and past.}}
\author[1]{Damiano Brigo}
\author[2]{Cristin Buescu}
\author[3]{Andrea Pallavicini}
\author[1]{Qing Liu}
\affil[1]{Department of Mathematics, King's College London, on move to Imperial College}
\affil[2]{Department of Mathematics, King's College London}
\affil[3]{Financial Engineering, Banca IMI, Milan}

\begin{document}

%
\date{First version: May 8, 2012. 
First posted on SSRN and arXiv on July 10, 2012.
\\
This version: 
\today
}

\maketitle

\begin{abstract}
We illustrate a problem in the self-financing condition used  in the papers ``Funding beyond discounting: collateral agreements and derivatives pricing" (\emph{Risk Magazine}, February 2010)  and ``Partial Differential Equation Representations of Derivatives with Counterparty Risk and Funding Costs" (\emph{The Journal of Credit
Risk}, 2011). 
These papers state an erroneous self-financing condition. In the first paper, this is equivalent to assuming that the equity position is self-financing on its own and without including
the cash position. In the second paper, this is equivalent to assuming that a subportfolio is self-financing on its own, rather than the whole portfolio. The error in the first paper is avoided when clearly distinguishing between price processes, dividend processes and gain processes. 
We present an outline of the derivation that yields the correct statement of the self-financing condition, clarifying
the structure of the relevant funding accounts, and show that the final result in ``Funding beyond discounting" is
correct, even if the self-financing condition stated is not.
\end{abstract}

\medskip

\noindent\textbf{AMS Classification Codes}: 91B70, 
 91G10, 
 91G20, 
 91G40  
\newline 
\textbf{JEL Classification Codes}: G12, G13 \newline

\noindent \textbf{Keywords}: {\small  Funding cost, cost of funding, funding and discounting, self-financing strategy, trading strategies, hedging.}

\pagestyle{myheadings} \markboth{}{{\footnotesize  D. Brigo, C. Buescu, A. Pallavicini, Q. Liu. Illustrating a problem in the self-financing condition}}

\section{Introduction}

Cost of funding has become a paramount topic in the industry. One just has to look at the number of presentations and streams at modeling conferences of 2010-2012 that deal with this topic to realize how much research effort is being put into it. Despite this sustained effort, the literature is still in its infancy. One of the difficulties is that the funding problem ties in naturally with credit risk, collateral modeling and discounting.

An initial analysis of the problem of replication of derivative transactions under collateralization in a Black and Scholes framework, but without the complicated and subtle intricacies of default risk, has been considered in \cite{Piterbarg2010}.

The fundamental impact of collateralization on default risk and on CVA and DVA has been analyzed in \cite{Cherubini}, and more recently in \cite{BrigoCapponiPallaviciniPapatheodorou} and \cite{BrigoCapponiPallavicini}.

The works \cite{BrigoCapponiPallaviciniPapatheodorou} and \cite{BrigoCapponiPallavicini} look at CVA and DVA gap risk under several collateralization strategies, with or without re-hypothecation, as a function of the margining frequency, with wrong-way risk and with possible instantaneous contagion. Minimum threshold amounts and minimum transfer amounts are also considered. We cite also \cite{Brigo2011} and \cite{BrigoMoriniPallavicini2012} for a list of frequently asked questions on the subject.

The fundamental funding implications in presence of default risk have been considered in \cite{MoriniPrampolini2011}, see also \cite{Castagna2011}.
In order to highlight some essential features of funding costs these works focus on particularly simple products, such as zero coupon bonds or loans.

The work in \cite{Fujii2010} analyzes implications of currency risk for collateral modeling.

The above references constitute a beginning for the funding cost literature, but do not have the level of generality needed to include all of the above features in a consistent framework that can then be used to manage complex products. A general theory is still missing. The only exceptions so far are \cite{BurgardKjaer2011a} and \cite{BurgardKjaer2011b}, which resort to a PDE approach and do not deal with the hidden complexities of collateral modeling and mark-to-market discontinuities at default, 
and do not state explicitly the funding assumptions for all the assets in their replicating portfolio.
The other general result is presented in the series \cite{Crepey2011}, \cite{Crepey2012a}, \cite{Crepey2012b}, which resort to backwards stochastic differential equations.

The paper \cite{Perini2011} follows the same level of generality of \cite{Crepey2011} and tries to deal with different models of funding policy, while still accounting for CVA, DVA, collateral and re-hypothecation.

A few available books have started including funding costs analysis, see for example \cite{kenyon} and \cite{BrigoMoriniPallavicini2012}.

In this note we address an important problem with the self-financing condition used in \cite{Piterbarg2010} and \cite{BurgardKjaer2011a}. The necessity for writing specifically a note is due to the fact that \cite{Piterbarg2010} is highly quoted at industry conferences worldwide in the practitioners space\footnote{
``For his work showing how to reconcile these complicating factors within a derivatives pricing framework
[the Author] wins Risk's quant of the year award for an unprecedented second time.
[The Author]'s paper ``Funding beyond discounting" was the first to show how funding rates and the posting of collateral can affect the pricing of derivatives."
\tt{http://www.risk.net/risk-magazine/feature/1934297/quant-vladimir-piterbarg-barclays-capital}}, 
while \cite{BurgardKjaer2011a} has received extensive exposure, and we believe it is therefore important to highlight this issue.

This erroneous self-financing condition stems from a common failure in applying the stochastic Leibnitz rule. This error is common even in mainstream textbooks such as \cite{hull} and has been pointed out explicity for example in \cite{shreve} (Exercise 4.10\footnote{We thank Chris Kenyon for reminding us of this exercise in \cite{shreve}, see also \cite{kenyon}}).   

%
%
%
%
%

\section{The self-financing condition and the problem in  \cite{Piterbarg2010}}

The problem is immediate.  We also notice that the same problems affects the proofs in \cite{BurgardKjaer2011a}.

Formula (2) in \cite{Piterbarg2010} reads, for the portfolio $\Pi$ that replicates the derivative $V$ (we use identical notation)

\[ V(t) = \Pi(t) = \Delta(t) S(t) + \gamma(t)  \ \ \ \hspace*{4cm}  (2)   \]
where $S$ is ``price process" of the underlying asset, and $\gamma$ is the ``cash amount split among a number of accounts [...]".

Then  \cite{Piterbarg2010} goes on:


``On the other hand, from (2), by the self-financing condition

\[ d \gamma(t) = d V(t) - \Delta(t)\ d S(t)  \ \ \ \hspace*{4cm} (*) \]

[...]"

We argue that this formulation of the self-financing condition is wrong.

It is enough to differentiate directly (2) to obtain

\[ d V(t) = d (\Delta(t) S(t)) + d \ \gamma(t),  \ \ \ \hspace*{4cm} (**)   \]

and combine this last equation with (*) to obtain

\[ \boxed{ d (\Delta(t) S(t)) = \Delta(t)\ d S(t) \ \ \ \mbox{(Wrong)} } . \]

The boxed equation, immediately implied  from the assumptions in \cite{Piterbarg2010}, would imply that the position in the risky asset $S$ is self-financing on its own, in that the change in the total value of the position, namely
$d (\Delta(t) S(t))$, is funded by the asset market movements alone: $\Delta(t)\ d S(t)$.

A further consequence of the above error follows immediately from the stochastic Leibnitz rule, leading to

\[ d \Delta_t = 0 \ \ \ \mbox{(wrong),}\]
and indeed if equity needs to be self-financing on its own the only possibility is that the amount of equity is constant (there is no re-balancing of the single position).

\section{The self-financing condition and the problem in \cite{BurgardKjaer2011a}}

The same problem affects the self-financing condition stated in \cite{BurgardKjaer2011a}.
In that work it is stated explicitly that the portfolio consisting of the stock $S$, the  bond $P_B$ of party $B$, the  bond $P_C$ of party $C$, and $\beta(t)$ cash is self-financing.
This portfolio value can be written as in the first Equation of \cite{BurgardKjaer2011a} following (3.2), namely (we use identical notation)
\[ - \hat{V}(t) = \Pi(t) = \delta(t) S(t) + \alpha_B(t) P_B(t) + \alpha_C(t) P_C(t) + \beta(t). \ \ \hspace*{3cm}  (\#) \]
The self-financing condition is stated in Equation (3.3) of \cite{BurgardKjaer2011a} which reads
\[ -  d \hat{V}(t) =  \delta(t) dS(t) + \alpha_B(t) dP_B(t) + \alpha_C(t) dP_C(t) + d \beta(t). \ \ \  \hspace*{3cm}(\# \#) \]
The reader has clearly seen where the problem is: if we now differentiate Equation $(\#)$ and equate the resulting equation with $(\# \#)$ we immediately obtain:
\[   d( \delta(t) S(t) + \alpha_B(t) P_B(t) + \alpha_C(t) P_C(t) ) =    \delta(t) dS(t) + \alpha_B(t) dP_B(t) + \alpha_C(t) dP_C(t),  \]
which clearly implies that the portfolio of the three assets
\[ S, \ P_B, \ P_C \]
is self-financing. This is clearly at odds with \cite{BurgardKjaer2011a} stating instead that the entire portfolio
\[ S, \ P_B, \ P_C, \ \mbox{cash} \]
is self-financing.

This is the same problem that afflicts the self-financing condition in \cite{Piterbarg2010}, except that here it is distributed across more than one asset.

\section{A sketch of the correct formulation in the framework of \cite{Piterbarg2010}}
Since the derivation of the result is important, as it provides a description of the funding account and of the funding strategy, we believe it is appropriate at this point to illustrate the proper formulation for \cite{Piterbarg2010}. We point out that we do not discuss the assumptions in \cite{Piterbarg2010}, as we rather concentrate on the correct formulation of the self-financing condition under their assumptions. For a more comprehensive framework that includes explicit default modeling, collateral modeling, re-hypothecation and debit valuation adjustments we refer the reader elsewhere, for example \cite{Perini2011} or \cite{Crepey2011}.

One of the problems in the above derivation is that it does not distinguish between gain processes, price processes and dividend processes, and in confusing them brings about a problem. We present below an informal and yet correct account of the theory. The full theory can be found in Duffie \cite{DuffieDAPT}.

We define an asset $A_i$ to be described by a price process $P_i^A$ and a dividend process $D_i^A$. The gain process is defined as
\[ G_i^A := P_i^A + D_i^A . \]
Now we assume that $i$ belongs to a finite set of indices $\{1,2,\dots, n\}$, and call $A$ the vector of assets with components $A_i$.

A trading strategy in $A$ is defined as a pair
\[ \Pi := (\theta^A,A) \]
where $\theta$ is a vector of portfolio positions (number of units) in each asset.

The 
value of the trading strategy is defined as
\[ V^\Pi := \theta_1^A P_1^A + \ldots + \theta_n^A P_n^A    \]
while the gain process of the strategy is
\[ d G^\Pi := \theta_1 d G^A_1 + \ldots +  \theta_n d G^A_n .  \]

A trading strategy is termed self-financing if its dividend is zero (there is no dividend outflow or inflow from the portfolio), namely if
\[ G^\Pi - V^\Pi = 0 . \]
This implies in particular that
\[ d G^\Pi = d V^\Pi  \]
or
\[ d (\theta_1^A P_1^A + \ldots + \theta_n^A P_n^A )  = \theta_1 d G^A_1 + \ldots +  \theta_n d G^A_n  \]
but notice that in general
\[ d (\theta_1^A P_1^A + \ldots + \theta_n^A P_n^A )  \neq \theta_1 d P^A_1 + \ldots +  \theta_n d P^A_n. \  \ \ \ \hspace*{3cm} (***) \]
Indeed, the self-financing condition for the strategy does not imply that
\[ d G_i^A = d P_i^A  \]
for the single asset $i$. Clearly, if all single assets have null dividend process, it follows that the self-financing condition for the strategy implies an equality in (***),
but more generally this does not hold.

Finally, a claim $Y$ is replicated by the strategy $\Pi$ if
\[ V^Y = V^\Pi, \ \  G^Y = G^\Pi \]
for all $t \ge 0$ and up to the claim maturity.

We now apply the above framework to the setup in \cite{Piterbarg2010}.

For a derivative with value $V^Y$ we seek a self-financing strategy that replicates it.
We try the strategy $(\theta^A,A)=:\Pi$.

We define the vector $A$ of assets:  $A_1$ to be a repo contract for the risky asset, the risky asset price being denoted by $S_t$;
$A_2$ is a collateral position, the collateral price being $C$, and $A_3$ is the funding position, whose price is denoted by $\alpha$.

For such positions we have the price processes

\begin{equation}\label{eq:dP}
P^A_1 = 0, \ \ P^A_2  = C, \ \ P^A_3 = \alpha
\end{equation}
and the gain processes
\begin{equation}\label{eq:dG}
 d G^A_1 = d S + (r_D - r_R) S dt , \ \ d G^A_2 = r_C C dt, \ \ d G^A_3 = r_F \alpha dt ,
\end{equation}
where $r_D$ is the rate at which stock dividends are paid, $r_R$ is the short rate on funding secured via repo,
$r_C$ is the short rate of cash/collateral, and $r_F$ is the short rate for unsecured funding.

We omit the superscript ``A" in the following to lighten notation.

One can immediately write the dividend processes for the three accounts as follows:
\begin{eqnarray}\label{eq:dD}
d D_1 &=& d G_1 - d P_1 = d S + (r_D - r_R) S dt, \ \ \ D_1(0) = 0,\\ \nonumber
d D_2 &=& d G_2 - d P_2 = r_C C dt - dC, \ \ \ D_2(0) = 0,\\  \nonumber
d D_3 &=& d G_3 - dP_3 = r_F \alpha dt - d \alpha , \ \ \  D_3(0) = 0 .
\end{eqnarray}

The components of $\theta$ are

\[ \theta_1 = \Delta, \ \ \theta_2 = 1, \ \ \theta_3 = 1 .\]


Notice that here the single asset dividend processes are not null. Hence we have (***).

The self-financing condition requires
\[ G^\Pi = V^\Pi \]
whereas the replication condition requires
\[  V^\Pi = V^Y .  \]
It follows that
\[ G^\Pi =  V^Y . \]
Let us now take a look at $V^\Pi$ and $G^\Pi$:
\[ V^\Pi = \Delta 0 + 1 C + 1 \alpha \ \Rightarrow  \alpha = V^\Pi - C = V^Y - C.  \]

We notice that, by plugging $\alpha = V^Y - C$ into Equations~(\ref{eq:dP}), (\ref{eq:dG}) and (\ref{eq:dD}), we obtain all the dynamics of the price, gain and dividend processes expressed in terms of the market observable quantities
\[ S, \ C, \ V  . \]

We have immediately by definition of gain process that
\[ d G^\Pi = \theta_1 d G_1 + \theta_2 d G_2+ \theta_3  d G_3,     \]
namely
\[ d G^\Pi = \Delta [ d S + (r_D-r_R) S  dt ]+ 1 r_C C dt + r_F (V^Y - C) dt.      \]

At the same time since we are imposing $G^\Pi =  V^Y$, we have $dV^Y = d G^\Pi$ given by the equation above. On the other hand, through Ito's formula,

\[ d V^Y = \frac{\partial V^Y}{\partial t} dt + \frac{\partial V^Y}{\partial S} dS + \frac{1}{2}\frac{\partial^2 V^Y}{\partial S^2} d \langle S \rangle^2  \]
and by equating the right hand sides of the last two equations we obtain

\[ \Delta [ d S + (r_D-r_R) S  dt ]+ 1 r_C C dt + r_F (V^Y - C) dt = \frac{\partial V^Y}{\partial t} dt + \frac{\partial V^Y}{\partial S} dS + \frac{1}{2}\frac{\partial^2 V^Y}{\partial S^2} d \langle S \rangle^2,  \]
from which

\[ \Delta = \frac{\partial V^Y}{\partial S}, \ \  \Delta (r_D-r_R) S   +  r_C C  + r_F (V^Y - C)  = \frac{\partial V^Y}{\partial t}   + \frac{1}{2}\frac{\partial^2 V^Y}{\partial S^2} \sigma_S^2 S^2. \]

From now on the derivation may continue as in \cite{Piterbarg2010} with their equation (3).

\section{Conclusion}
We illustrated a problem with the self-financing condition used in \cite{Piterbarg2010} and \cite{BurgardKjaer2011a}. This wrong condition implies that the
portfolio without cash 
is self-financing on its own. Since a sound funding theory depends crucially on financing costs, the violation of the self-financing condition is an important problem. Fortunately, the error can be corrected and the same result obtained in a rigorous way for \cite{Piterbarg2010}, and we briefly sketched the correct setup, casting light on the necessary distinction between dividend processes, price processes and gain processes.

%




\begin{thebibliography}{10}

\bibitem{Brigo2011}
D.~Brigo.
\newblock Counterparty risk {F}{A}{Q}: Credit {V}a{R}, {P}{F}{E}, {C}{V}{A},
  {D}{V}{A}, closeout, netting, collateral, re-hypothecation, {W}{W}{R},
  {B}asel, funding, {C}{C}{D}{S} and margin lending.
\newblock {\em Working Paper}, 2011.

\bibitem{BrigoCapponiPallavicini}
D.~Brigo, A.~Capponi, and A.~Pallavicini.
\newblock Arbitrage-free bilateral counterparty risk valuation under
  collateralization and re-hypothecation with application to {CDS}.
\newblock {\em Mathematical Finance}, 2011.
\newblock Accepted for publication.

\bibitem{BrigoCapponiPallaviciniPapatheodorou}
D.~Brigo, A.~Capponi, A.~Pallavicini, and V.~Papatheodorou.
\newblock Collateral margining in arbitrage-free counterparty valuation
  adjustment including re-hypotecation and netting.
\newblock {\em Working Paper}, 2011.

\bibitem{BrigoMoriniPallavicini2012}
D.~Brigo, M.~Morini, and A.~Pallavicini.
\newblock {\em Counterparty Credit Risk, Collateral and Funding with pricing
  cases for all asset classes}.
\newblock Wiley, Forthcoming, 2012.

\bibitem{BurgardKjaer2011b}
C.~Burgard and M.~Kjaer.
\newblock In the balance.
\newblock {\em Risk Magazine}, October, 2011.

\bibitem{BurgardKjaer2011a}
C.~Burgard and M.~Kjaer.
\newblock Partial differential equation representations of derivatives with
  counterparty risk and funding costs.
\newblock {\em The Journal of Credit Risk}, 7 (3):1--19, 2011.

\bibitem{Castagna2011}
A.~Castagna.
\newblock Funding, liquidity, credit and counterparty risk: Links and
  implications.
\newblock {\em Working Paper}, 2011.

\bibitem{Cherubini}
U.~Cherubini.
\newblock Counterparty risk in derivatives and collateral policies: the
  replicating portfolio approach.
\newblock In L.~Tilman, editor, {\em ALM of Financial Institutions}.
  Institutional Investor Books, 2005.

\bibitem{Crepey2011}
S.~Cr\'epey.
\newblock A {BSDE} approach to counterparty risk under funding constraints.
\newblock {\em Working Paper}, 2011.

\bibitem{Crepey2012a}
S.~Cr\'epey.
\newblock Bilateral counterparty risk under funding constraints – {P}art {I}:
  {P}ricing.
\newblock {\em Forthcoming in Mathematical Finance}, 2012.

\bibitem{Crepey2012b}
S.~Cr\'epey.
\newblock Bilateral counterparty risk under funding constraints – {P}art
  {II}: {CVA}.
\newblock {\em Forthcoming in Mathematical Finance}, 2012.

\bibitem{bibscandals}
K.~Crowley and A.~Choudhury.
\newblock Bloomberg.
\newblock {\em
  http://www.bloomberg.com/news/2012-07-
  
  05/made-in-london-scandals-risk-city-s%
-reputation-as-finance-center.html}, 6 July 2012.

\bibitem{DuffieDAPT}
D.~Duffie.
\newblock {\em Dynamic Asset Pricing Theory}.
\newblock Princeton University Press, 3rd edition, 2001.

\bibitem{Fujii2010}
M.~Fujii, Y.~Shimada, and A.~Takahashi.
\newblock Collateral posting and choice of collateral currency.
\newblock {\em Working Paper}, 2010.

\bibitem{hull} J.~Hull. 
\newblock Options, Futures and other Derivatives. 8th Edition. 
\newblock Prentice Hall, 2011.

\bibitem{kenyon}
C.~Kenyon, and R.~Stamm. 
Discounting, Libor, CVA and Funding. 
\newblock Palgrave MacMillan, 2012.

\bibitem{MoriniPrampolini2011}
M.~Morini and A.~Prampolini.
\newblock Risky funding: A unified framework for counterparty and liquidity
  charges.
\newblock {\em Risk Magazine}, March, 2011.

\bibitem{Perini2011}
A.~Pallavicini, D.~Perini, and D.~Brigo.
\newblock Funding {V}aluation {A}djustment: {FVA} consistent with {CVA}, {DVA},
  {WWR}, {C}ollateral, {N}etting and re-hyphotecation.
\newblock {\em 
ssrn.com}, 2011.

\bibitem{Piterbarg2010}
V.~Piterbarg.
\newblock Funding beyond discounting: collateral agreements and derivatives
  pricing.
\newblock {\em Risk Magazine}, 2:97--102, 2010.

\bibitem{shreve} S.~Shreve. 
\newblock Stochastic Calculus for Finance II: Continuous-Time Models. 
\newblock Springer Verlag, Heidelberg, 2004.


\end{thebibliography}

\end{document}